\documentclass[aip,jcp,reprint]{revtex4-1}
\usepackage{epsfig}
\usepackage{graphicx}
\usepackage{amsmath}
\usepackage{amssymb}
\usepackage{amsfonts}  
\usepackage{fancyhdr}
\usepackage{bm}
\usepackage{dcolumn}
\usepackage{subfigure}
\newcommand{\bra}[1]{\langle #1|}
\newcommand{\ket}[1]{|#1\rangle}
\newcommand{\braket}[2]{\langle #1|#2\rangle}

\newcommand{\half}{\frac{1}{2}}

\newcommand{\proj}{\mathcal{P}}

\begin{document}
\title{Flux-correlation approach to characterizing reaction pathways in quantum systems: 
A study of condensed-phase proton-coupled electron transfer}
\author{Nandini Ananth}
\author{Thomas F. Miller, III}
\affiliation{Division of Chemistry and Chemical Engineering, California Institute of Technology, Pasadena, California 91125, USA} %
\begin{abstract}
We introduce a simple method for characterizing reactive pathways in quantum systems.  Flux auto-correlation and cross-correlation functions are employed to develop a quantitative measure of dynamical coupling  
in quantum transition events, such as reactive tunneling and resonant energy transfer.  We utilize the  method to study condensed-phase proton-coupled electron transfer (PCET) reactions and to determine the relative importance of competing concerted and sequential reaction pathways.  Results presented here include numerically exact quantum dynamics simulations for model condensed-phase PCET reactions.  This work demonstrates the applicability of the new method for the analysis of both approximate and exact  quantum dynamics simulations.\\\\
\end{abstract}
\maketitle
\section{Introduction}
Detailed mechanistic understanding of condensed-phase chemical 
dynamics is essential for the design of next-generation 
molecular catalysts and photosystems.
The role of numerical simulations 
in this effort is particularly important for 
systems that exhibit multiple, competing 
reactive pathways.
The characterization of reactive pathways
in classical systems has been greatly 
advanced by the development of methods for
rare event sampling \cite{pgb02,gh04,tsve05,tfm07a,tfm07b,ew07}.
However, the corresponding tools for quantum systems 
\cite{shn80,am03,ph10,jyz11,nb11} are less developed,
despite significant demand for understanding reactive 
charge and energy transfer pathways in complex quantum systems.

In this paper, we introduce a  method 
for 
characterizing reactive pathways and 
quantifying dynamical correlations in 
general quantum systems using real-time 
flux auto-correlation (FAC) and flux 
cross-correlation (FCC) functions.
We employ the new approach to investigate the reaction dynamics of condensed-phase proton-coupled
electron transfer (PCET), an important  prototype 
for quantum systems that exhibit multiple reaction 
pathways \cite{ric98,shs01,jmm04,mhh07,vrik11,jb11,avs11}.
Numerical results demonstrate the practicality and utility of the new 
method for the mechanistic analysis of coupled quantum dynamical processes.

\section{Theory}
We consider a set of states $\{\ket{n}\}$ that span
the full Hilbert space of a closed quantum system,
and we consider the partitioning of this set of states
into $N$ subsets $\{\Omega_j\}$, such that each
state belongs to one and only one subset.
The criteria for this partitioning are flexible; for instance, 
cut-offs based on energy expectation values can be used to 
generate subsets of states with similar energies, 
and conditions based on position expectation values can be used to generate 
subsets of states with similar configurations.
The dividing surface corresponding to subset
$\Omega_j$ 
is 
defined in terms of the projection 
operator
\begin{equation}
\proj_j =\sum_{i=1}^{\text{all states}}\ket{n_i}\bra{n_i}\epsilon_{ij},
\label{eq:proj_def}
\end{equation}
where $j=1,\ldots N$, and 
\begin{equation}
\epsilon_{ij}=\left\{
\begin{matrix}
1 & ,\;\;\;\ket{n_i}\in \Omega_j\\
0 & ,\;\;\;\ket{n_i}\notin \Omega_j
\end{matrix}
\right. .
\label{eq:set_occ}
\end{equation}

The net flux associated with transitions from subset 
$\Omega_j$ to the remainder of the Hilbert space
is given by
\begin{equation}
   F_j=\frac{i}{\hbar}\left[ H, \proj_j\right],
   \label{eq:flux_def}
\end{equation}
where $H$ is the Hamiltonian for the full system.
Dynamical correlations between these transitions 
can be obtained
from thermal, real-time FAC and FCC functions
 \cite{whm74,shn80,whm83,gav89,tcg97}, 
\begin{equation}
   C_{mj}(t)=\text{Tr}\left[ e^{-\beta H} F_m e^{iHt} F_j
   e^{-iHt} \right],
   \label{eq:fcc_def}
\end{equation}
where $\beta$ is the reciprocal temperature, and 
the FAC function corresponds to the case $m~=~j$. 
Physically, Eq. (\ref{eq:fcc_def}) describes the correlation
between transitions into/out-of subset $\Omega_m$ with
transitions into/out-of subset $\Omega_j$.

Using that the net flux for a closed quantum
system is zero, i.e. $\sum_{j=1}^NF_j~=~0$, the FAC
function can be expressed as a linear combination of 
FCC functions; specifically,
\begin{equation}
 C_{jj}(t)=-\sideset{}{'}\sum_{m=1}^{N} C_{mj}(t),
 \label{eq:fac_fcc}
\end{equation}
where the summation excludes the $m=j$ term.
Introducing the zeroth moments of the real part of these correlation functions,
\begin{equation}
	k_{mj}=\int_0^\infty dt\;\text{Re}[C_{mj}(t)],
\label{eq:def_notrate}
\end{equation} 
Eq. (\ref{eq:fac_fcc}) then yields
\begin{equation}
k_{jj} = -\sideset{}{'}\sum_{m=1}^{N} k_{mj}. 
\label{eq:notrate_decomp}
\end{equation}

Within the assumption that $\proj_j$ divides
the system into two basins of stability,
Eq. (\ref{eq:notrate_decomp}) provides a
decomposition of the corresponding reaction rate.
We thus introduce
 a measure of the degree
to which dynamical correlations between 
transitions associated with subsets $\Omega_j$ and 
$\Omega_m$ contribute to the overall reaction
rate,
\begin{equation}
	\kappa_{mj} = -\frac{k_{mj}}{k_{jj}}.
	\label{eq:dcf_def}
\end{equation}
More generally, regardless of metastability of
subsets, the dynamical correlation factor (DCF) defined 
in Eq. (\ref{eq:dcf_def}) provides a 
transferable 
measure of the relative contribution  
from subset $\Omega_m$ to the transient dynamics 
associated with entering/leaving subset $\Omega_j$. 
Non-zero values for the DCF 
indicate correlated transitions between the subsets and provide a basis for identifying important dynamical pathways and reaction mechanisms.
We note that the DCF can be calculated using either exact or approximate quantum dynamical methods \cite{na10}, and 
flexibility in the definition of subset partitions enables the detailed characterization of 
complex systems.
Furthermore, to characterize processes that  are far from thermal equilibrium, 
DCFs can be similarly  constructed in terms of 
flux correlation functions with non-Boltzmann initial distributions.

\section{PCET model systems}
To demonstrate this flux-correlation
approach for the analysis of systems with 
competing dynamical pathways, 
we consider a series of condensed-phase PCET reactions. 
The reactions are described using a system-bath Hamiltonian,
\begin{eqnarray}
	\nonumber
	H &=& \frac{p_s^2}{2m_s}+\frac{p_x^2}{2m_x}+V_\text{p}(x)
	+ V_\text{ps}(x,s) + V_e(x,s) + \\
	&&+\sum_j \frac{P_j^2}{2M_j} + 
	\half M_j\omega_j^2(Q_j - \frac{c_js}{M_j\omega_j^2})^2,
	\label{eq:pcet_ham}
\end{eqnarray}
where the 
coordinates $x$, $s$ and $Q_j$ 
correspond to the proton, solvent polarization, and 
bath modes with masses of $m_x$, $m_s$, and $M_j$, respectively.
Potential energy surfaces associated with the donor and 
acceptor 
electronic 
states are described in the 
diabatic representation,
\begin{equation}
V_e(x,s)=
	\left(
	\begin{array}{cc}
		V_{11}(s)+V_\text{ep}(x) & V_{12}(s) \\
		V_{12}(s) & V_{22}(s)-V_\text{ep}(x)
	\end{array}
	\right),
\label{eq:bare_et}
\end{equation}
where 
$V_{ii}(s)~=~\half m_s\omega_s^2(s~-~s_i)^2$ with $i=1,2$,
the electron-proton coupling is given by $V_\text{ep}(x)~=~\mu_2\,\text{tanh}(\phi x)$,
and $V_{12}(s)$ is the diabatic coupling.

The term $V_\text{p}(x)$ in Eq. (\ref{eq:pcet_ham}) describes a double-well potential for
the proton coordinate, 
\begin{equation}
	V_\text{p}(x)=-\frac{m_x\omega_x^2}{2}x^2 +
	\frac{m_x^2\omega_x^4}{16V_0}x^4-\lambda x^3,
	\label{eq:bare_pt}
\end{equation}
where $\omega_x$ is the proton vibrational frequency, and
$V_0$ is the intrinsic proton transfer barrier height. 
The proton solvent interaction is described using
$V_\text{ps}(x,s)~=~-~\mu_1 s\,\text{tanh}(\phi x)$.

The dissipative bath in Eq. (\ref{eq:pcet_ham}) exhibits an 
Ohmic spectral density,
\begin{equation}
	J(\omega)=\eta\omega e^{-\omega/\omega_c},
\end{equation}
where $\omega_c$ is the cut-off frequency, and 
$\eta$ is the dimensionless coupling between solvent 
and bath modes \cite{aoc83}.
The spectral density is discretized using 
$f$ oscillators with frequencies \cite{irc05}
\begin{equation}
	\omega_j=-\omega_c\,\text{log}\left( \frac{j-0.5}{f} \right)
\end{equation}
and coupling constants
\begin{equation}
	c_j = \omega_j
	\left( \frac{2\eta M_j\omega_c}{f \pi} \right),
\end{equation}
where $j=1,\ldots f$.

PCET reactions proceed via concerted or sequential 
mechanisms, depending on the chronology of the electron 
transfer (ET) and proton transfer (PT) events.
The concerted mechanism involves dynamically correlated
transfer of both the electron and proton, whereas 
the sequential mechanism involves dynamically uncorrelated
ET and PT steps.
We consider three models for PCET in the current study, each of which corresponds to a different set of parameters for the Hamiltonian in Eq.\@ (\ref{eq:pcet_ham}). 
Models I and II are adapted from an earlier study of uni-directional
PCET in non-dissipative systems 
and assume that the solvent responds to the transferring electron and proton as if they  are like-charged
particles \cite{jyf97}. 
This unphysical picture 
of  solvent polarization is corrected in model III, 
in which the solvent response to the transferring electron is 
counteracted by the solvent response to the transferring proton. 
The parameters for all three models are provided in 
Table~{\ref{table:model_params}}.

\begin{table}[!htb]
   \caption{Parameters for the model PCET systems.}
	\begin{tabular}{cccc}\toprule
		\multicolumn{1}{c}{Parameter$^{\rm a}$}
		&\multicolumn{1}{c}{Model I}
& \multicolumn{1}{c}{Model II}
& \multicolumn{1}{c}{Model III}
\\
   \hline
    $m_x$ & 1836.1 & 1836.1 & 1836.1 \\
    $\omega_x$ & 0.0104 & 0.0104 & 0.0104\\
    $V_0$ & 0.012  & 0.014 & 0.012\\
    $m_s$ & 22000 & 22000 & 22000 \\
    $\omega_s\times10^{4}$ & 3.72 & 4.00 & 3.72\\
    $s_1$ & -2.13  & -2.16 & -2.13\\
    $s_2$ & 2.13 & 2.16 & 2.13\\
    $V_{12}$ & 0.0245 & 0.0124 & 0.0245 \\
    $\mu_1$ & 0.0011 & 0.017 & -0.0011\\
    $\mu_2\times10^3$ & 5.84 & 0.71 & 5.84\\
    $\phi$ & 8.0 & 8.0 & 8.0 \\
    $\lambda$ & 0.0  & 0.012 & 0.0\\
    $f$ & 12 & 12 & 12 \\
    $M_j$ & $m_s$ & $m_s$ & $m_s$\\
    $\eta$ & $m_s\omega_s$ & $m_s\omega_s$ & $m_s \omega_s$\\
    $T$/K & $300$ & $750$ & $300$\\
    \botrule
\end{tabular}
\footnotetext{All values in atomic units, unless otherwise specified.}
   \label{table:model_params}
\end{table}
\section{Calculation Details }
For each PCET model system, 
FAC and FCC functions are calculated
using the quasi-adiabatic path-integral (QUAPI) method
 \cite{nm92,dem93,mt93a,mt93b,mt96}. 
QUAPI is a numerically exact quantum dynamics method that
employs a real-time path integral formulation. 
It has previously been used to study single-particle transfer reactions, such as ET or PT \cite{mt93b,mt96}; here, we extend the approach
to describe the coupled transfer of both an electron and a proton. 
Details of the QUAPI implementation are provided in Appendix I.

We partition the full set of quantum states
for the PCET Hamiltonian (Eq. (\ref{eq:pcet_ham})) into $N=3$ 
subsets that are defined in terms of the proton position and the electronic
diabatic state. 
The subset associated with the PCET reactant states, 
 $\Omega_\text{r}$, is defined in terms of the projection operator
 \begin{equation}
\proj_\text{r}=\ket{1}\bra{1}h(-x).
\label{eq:react_proj}
\end{equation}
Similarly, $\Omega_\text{p}$  includes the PCET product states and is defined using
 \begin{equation}
\proj_\text{p}=\ket{2}\bra{2}h(x),
\label{eq:prod_proj}
\end{equation}
and 
$\Omega_\text{i}$ includes intermediate states associated with  single-particle transfer and is defined using
 \begin{equation}
\proj_\text{i}=\ket{1}\bra{1}h(x)+
\ket{2}\bra{2}h(-x).
\label{eq:int_proj}
\end{equation}
In these equations,  $\ket{1}$ and $\ket{2}$ indicate
the donor and acceptor electronic diabatic states,
and  $h(x)$ is the heaviside function 
\begin{equation}
 h(x)=\left\{ \begin{matrix}
 1 & ,\;\;\;x>0\\
 0 & ,\;\;\;x<0.
 \end{matrix}\right.
\label{eq:hside_def}
\end{equation}
 Using these subset definitions, the corresponding FAC and FCC functions 
  are calculated using Eq.\@ (\ref{eq:fcc_def}). 
  Finally, Eq.\@ (\ref{eq:dcf_def}) is used to calculate $\kappa_{\textrm{ip}}$ and $\kappa_{\textrm{rp}}$, which quantify 
  dynamical correlations of the product subset with the intermediate and reactant subsets, respectively.   
 For the present case in which the states are simply partitioned into  reactant, product, and intermediate subsets, we can  relate the calculated DCF to competing reaction mechanisms.
 Specifically, by reporting on whether the reaction dynamics proceeds via the intermediate subset, or whether it proceeds via direct transfer from the reactant to product subsets,    $\kappa_{\textrm{ip}}$ and $\kappa_{\textrm{rp}}$ respectively indicate the 
 degree to which the sequential or concerted PCET mechanism is dominant. 

\section{Results}
\begin{figure}[!htb]
	\includegraphics[angle=0,scale=0.35]{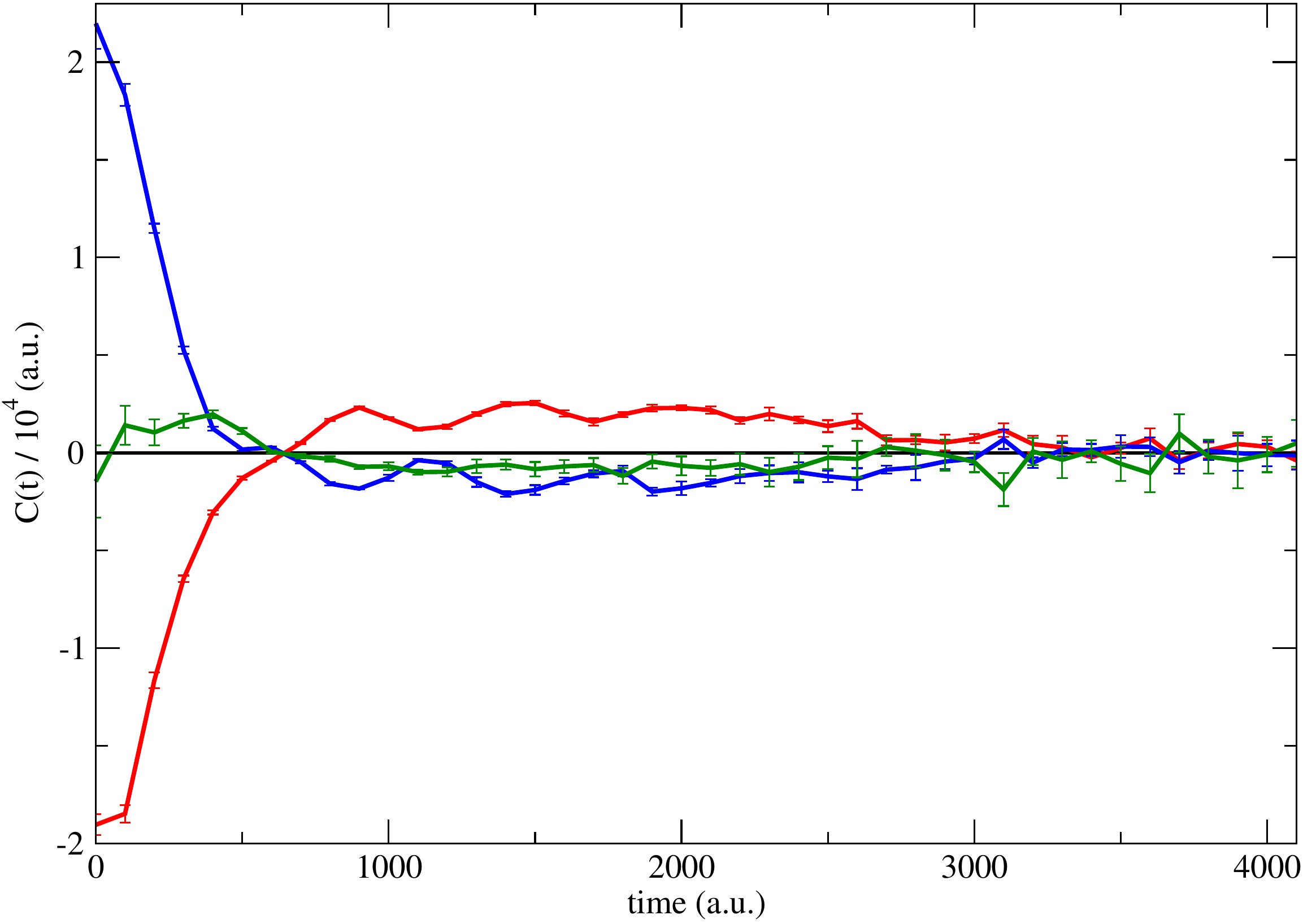}
	\caption{
	Numerically exact, symmetrized flux-correlation functions for model I,
	with $C_\textrm{pp}(t)$ in blue, $C_\textrm{rp}(t)$ in red, 
	and $C_\textrm{ip}(t)$ in green.
  }
  \label{fig:model1_fcc}
  \vspace{0cm}
\end{figure}

Fig. \ref{fig:model1_fcc} presents the calculated FAC and FCC functions for model I, with $C_\text{pp}(t)$, $C_\text{rp}(t)$, and $C_\text{ip}(t)$ plotted in blue, red, and green, respectively.
The FAC function exhibits initial decay on the 500 a.u. timescale, followed by modest recrossing.
Cross-correlations in the subset dynamics are most pronounced for the reactant and product subsets, with  $C_\text{rp}(t)$ closely mirroring the features of the product FAC function.
In contrast, only a small degree of cross-correlation is found for the dynamics associated with the intermediate and product subsets.
All dynamical correlations among the subsets are found to vanish by approximately 4000 a.u. in time.
\vspace{-0.1cm}

Integration of these correlation functions yields the DCF for model I, 
which are reported in Table \ref{table:model1_dcf}.
The larger magnitude of $\kappa_{\textrm{rp}}$ indicates that the PCET reaction in model I primarily proceeds via the concerted mechanism, 
although a substantial contribution from the sequential pathway is also found.
We note that earlier simulations for a non-dissipative version 
of this PCET model  also concluded the importance of the concerted mechanism \cite{jyf97,ss00}.

\begin{table}[!htb]
\caption{Reaction mechanisms for PCET in models I-III.}
\begin{tabular}{ccccc}\toprule
\multicolumn{1}{c}{Pathway}
&\multicolumn{1}{c}{DCF}
&\multicolumn{1}{c}{Model I}
&\multicolumn{1}{c}{Model II}
&\multicolumn{1}{c}{Model III}
\\ \hline
Concerted & $\kappa_{\textrm{rp}}$ & 0.62(6) & 0.010(7) & 0.93(7) \\
Sequential& $\kappa_{\textrm{ip}}$ & 0.38(5) & 0.99(5) & 0.07(3) \\
\botrule
\end{tabular}
\label{table:model1_dcf}
\end{table}

\begin{figure}[!htb]
	\includegraphics[angle=0,scale=0.35]{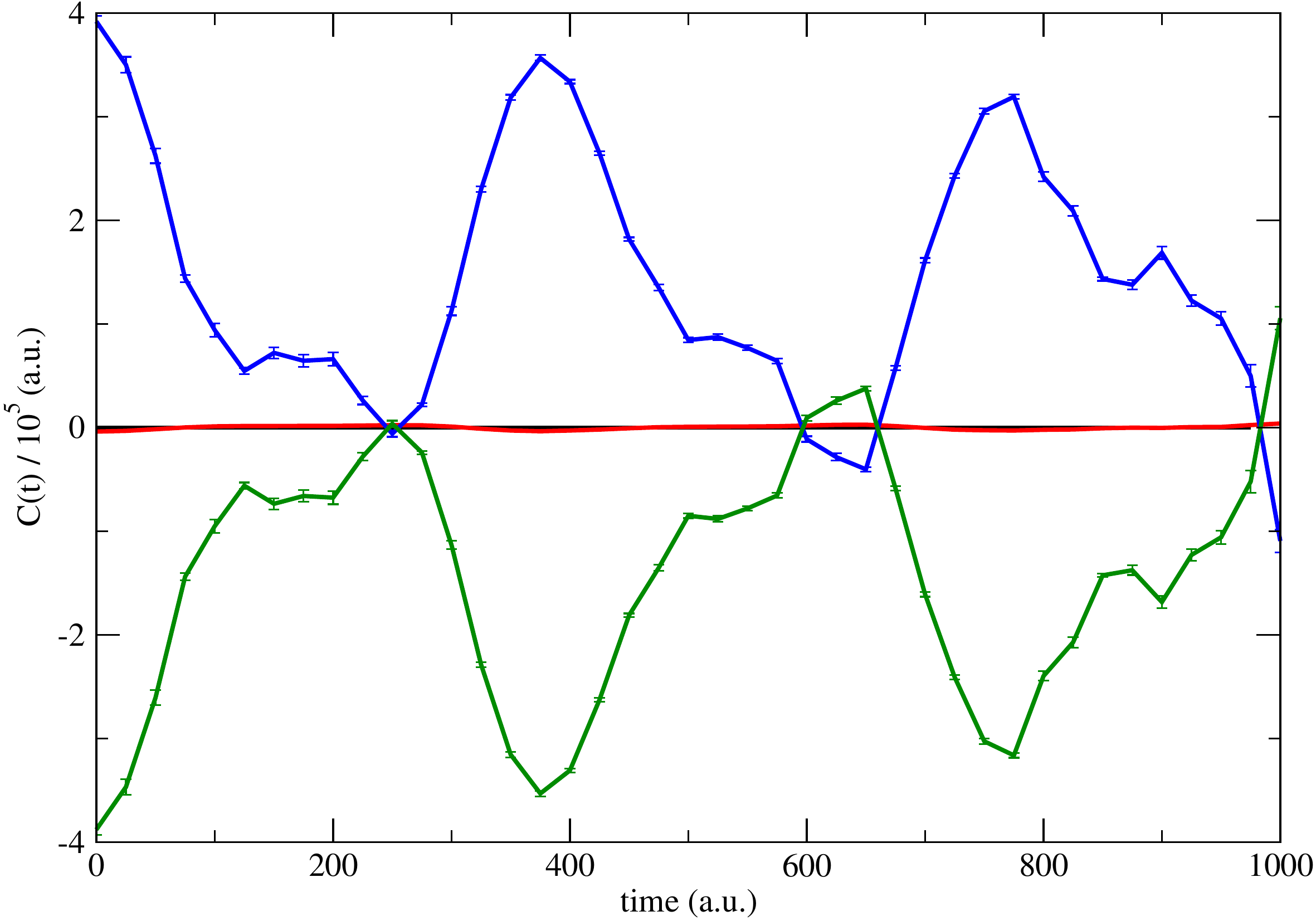}
	\caption{
	Numerically exact, symmetrized flux-correlation functions for model II,
	with $C_\textrm{pp}(t)$ in blue, $C_\textrm{rp}(t)$ in red, 
	and $C_\textrm{ip}(t)$ in green. 
	}
  \label{fig:model2_fcc}
  \vspace{0cm}
\end{figure}

Fig.\@ \ref{fig:model2_fcc} presents the $C_\text{pp}(t)$, $C_\text{rp}(t)$, and $C_\text{ip}(t)$ correlation functions 
for model II at short times,
which reveal significant differences in comparison to 
Fig. \ref{fig:model1_fcc}.
The FAC function 
for model II exhibits pronounced oscillations 
on the timescale of proton vibrations,
but  more striking differences are seen in the 
cross-correlation functions. 
In Fig.\@ \ref{fig:model2_fcc}, far greater contributions are seen from 
the $C_{\textrm{ip}}$ than $C_{\textrm{rp}}$, indicating that flux into the product state is dynamically coupled with the single-particle transfer intermediates, rather than with the reactant subset.
This point is further illustrated 
in Table \ref{table:model1_dcf}, which reveals $\kappa_\textrm{ip}$ to approach 
unity while $\kappa_\textrm{rp}$ nearly vanishes. 
These results clearly indicate that model II exhibits a sequential PCET reaction mechanism, 
in agreement with   studies of a related, non-dissipative system \cite{jyf97,ss00}.
\vspace{-0.1cm}

\begin{figure}[!htb]
	\includegraphics[angle=0,scale=0.35]{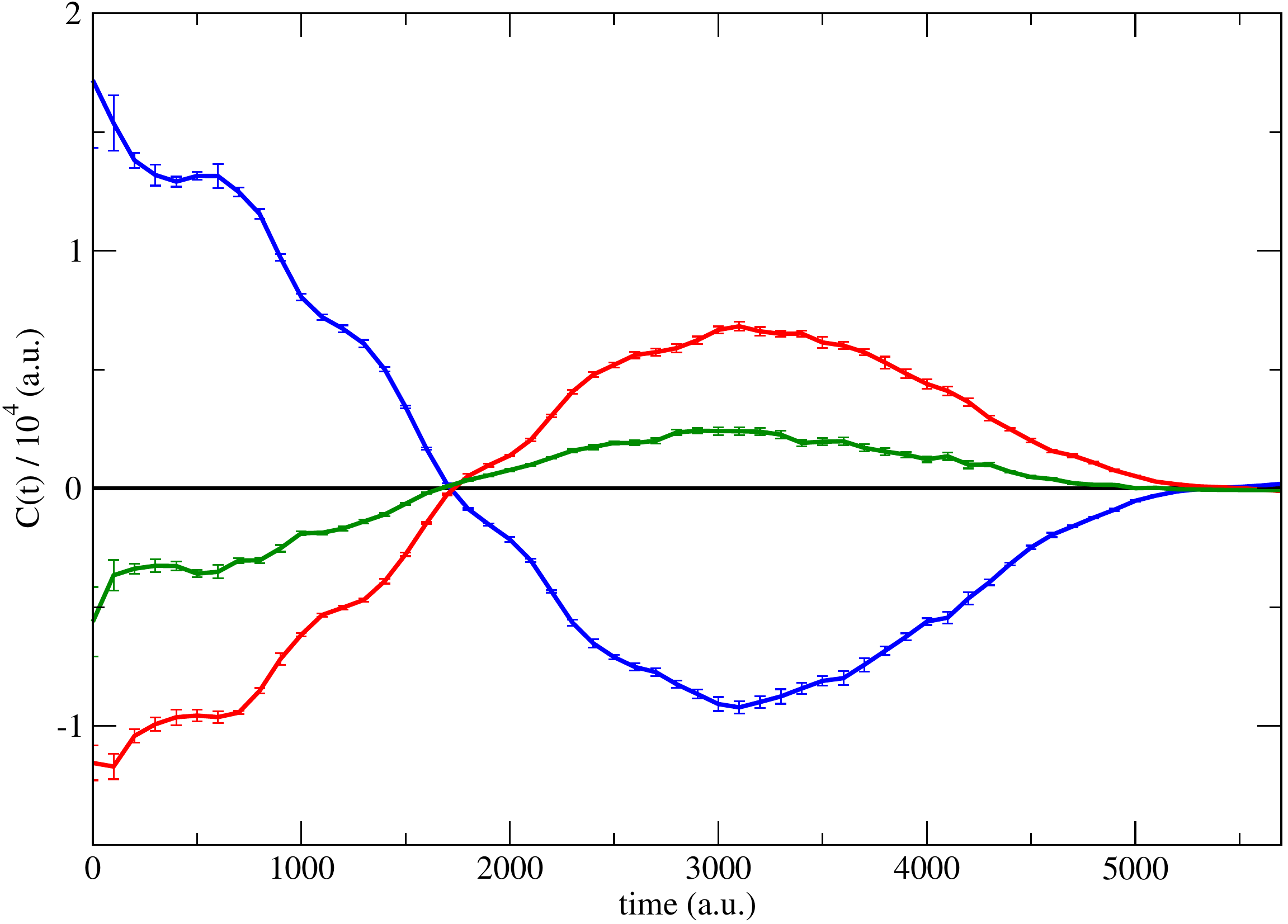}
	\caption{
	Numerically exact, symmetrized flux-correlation functions for model III,
	with $C_\textrm{pp}(t)$ in blue, $C_\textrm{rp}(t)$ in red, 
	and $C_\textrm{ip}(t)$ in green. 
  }
  \label{fig:model3_fcc}
  \vspace{0cm}
\end{figure}

Finally,  Fig.\@ \ref{fig:model3_fcc} presents the  
flux-correlation functions for  model III,
which differs from model I only in terms of the solvent polarization response to the proton coordinate  (Table~{\ref{table:model_params}}). 
The 
FAC function for model III 
exhibits a slower timescale for initial decay than for model I, as well as a more pronounced degree of  dynamical recrossing.  
Both FCC functions 
reflect the time dependence of the FAC function, although the magnitude of $C_\textrm{rp}(t)$ is greater than that of $C_\textrm{ip}$
at all times.
Interestingly, Table~\ref{table:model1_dcf} shows that $\kappa_\textrm{rp}$ 
approaches unity, whereas $\kappa_\textrm{ip}$ nearly vanishes, an
indication that 
the PCET reaction in model III is more dominated by the concerted 
mechanism than the corresponding concerted reaction in model I. 
In model III, the solvent response to the net-neutral PCET charge-transfer reaction yields reactant and product states with less solvent polarization than in model I; 
the energetic favorability of minimizing solvent reorganization throughout the reaction thus creates a driving force for co-localizing the charge distributions of the transferring electron and proton, which leads to the more strongly concerted PCET mechanism observed for model III.

\vspace{-0.1cm}

\section{Conclusions}
We introduce a method in which flux-correlation functions are used to characterize dynamical correlations and reaction pathways in quantum systems.
Numerical results demonstrate the utility of the method for the analysis of exact quantum dynamic simulations of 
condensed-phase PCET.
In particular, we 
quantify the relative importance of the sequential and concerted PCET reaction mechanisms for two previously studied model systems, 
and in a third  system, we demonstrate that introduction of a refined description for the solvent-proton coupling leads to greater dominance of the 
concerted PCET mechanism. 
It is expected that this approach will prove useful in the future analysis of reaction and energy transfer pathways in complex  systems.

\section*{Acknowledgements}
This work was supported by the  
U.S.~Department of Energy (DOE), Chemical Sciences, Geosciences and Biosciences Division, Office of Basic Energy Sciences under Grant No. DE-FG02-11ER16247
and the
National Science Foundation (NSF) CAREER Award under Grant No.~CHE-1057112. 
Additionally, T.F.M.~acknowledges support from an Alfred P.~Sloan Foundation Research Fellowship.
Computational resources were provided by the National Energy Research
Scientific Computing Center, which is supported by the Office of
Science of the U.S. Department of Energy under Contract No. 
DE-AC02-05CH11231.

\section*{Appendix I: QUAPI implementation for PCET}
\label{sec:quapi}
Here, we describe the implementation of the QUAPI method used
in the current study.
The total PCET Hamiltonian in Eq. (\ref{eq:pcet_ham}) is written as a sum
of system and bath contributions such that, $H=H_s+H_B$, where
\begin{equation}
\label{eq:sys_hamiltonian}
	H_\mathrm{s} = \frac{p_s^2}{2m_s}+\frac{p_x^2}{2m_x}+V_\text{e}(x,s)+
	V_\text{p}(x)+V_\text{ps}(x,s)
\end{equation}
and the Hamiltonian describing the bath modes and their coupling to the solvent coordinate is
\begin{equation}
	H_\mathrm{B}= \sum_{j=1}^f \frac{P_j^2}{2M_j} + \sum_{j=1}^f\half M_j\omega_j^2
   \left(Q_j-\frac{c_js}{M_j\omega_j^2}\right)^2.
   \label{eq:quapi_bath_hamiltonian}
\end{equation}
In practice, we report the symmetrized form of the flux-flux correlation functions
and note that the zeroth moment of these functions is identical to that of the
standard correlation function defined in Eq. (\ref{eq:fcc_def}) \cite{whm83}.
The symmetrized FCC function is given by
\begin{equation}
	C_{jm}(t) = \mathrm{Tr}[F_j e^{iHt_{\mathrm{c}}^{*}/\hbar}F_m e^{-iHt_{\mathrm{c}}/\hbar}], 
	\label{eq:cff_defn}
\end{equation}
where the subset indices $j,m\in\{\text{i},\text{r},\text{p}\}$, and ${t_{\mathrm{c}}=t-i\beta\hbar/2}$. 
The complex-time propagators in Eq. (\ref{eq:cff_defn}) are discretized 
into $\mathcal{N}$ time slices 
of length $\Delta t_{\mathrm{c}}$, and
the trace over bath modes is evaluated analytically to yield
\begin{eqnarray}
\label{eq:cdef}
       \nonumber
       C_{jm}(t)&=& \int d\mathbf{s} \int d\mathbf{x} 
              \;\sum_{\boldsymbol{\sigma}=1}^2 \;
	\mathcal{I}(\mathbf{s})\;K(\mathbf{s},\mathbf{x},\boldsymbol{\sigma};t_{\mathrm{c}})\\
	\times &&  \bra{s_1,x_1,\sigma_1}F_j\ket{s_{2\mathcal{N}+2},x_{2\mathcal{N}+2},\sigma_{2\mathcal{N}+2}} \\
	\nonumber
	 \times && \bra{s_{\mathcal{N}+2},x_{\mathcal{N}+2},\sigma_{\mathcal{N}+2}}F_m\ket{s_{\mathcal{N}+1},x_{\mathcal{N}+1},\sigma_{\mathcal{N}+1}},
\end{eqnarray}
where ${\mathbf{s}=\{s_1, \ldots, s_{2\mathcal{N}+2}\}}$,
 ${\mathbf{x}=\{x_1, \ldots, x_{2\mathcal{N}+2}\}}$, and ${\boldsymbol{\sigma}=\{\sigma_1,\ldots,\sigma_{2\mathcal{N}+2}\}}$. Numerical evaluation of the integrals in Eq. (\ref{eq:cdef}) 
 is performed using two independent path-integral Monte Carlo simulations (MC1 and MC2), 
as we have previously described in detail \cite{arm11}.

The term $K(\mathbf{s},\mathbf{x},\boldsymbol{\sigma};t_\mathrm{c})$
in Eq.~\ref{eq:cdef} is the path-integral representation for the 
complex-time propagators of the system Hamiltonian,  
\begin{eqnarray}
\label{eq:k_prod}
        \nonumber
K(\mathbf{s},&\mathbf{x},&\boldsymbol{\sigma};t_{\mathrm{c}})=
\prod_{k=2}^{\mathcal{N}+1}\bra{s_k,x_k,\sigma_k}e^{-iH_\mathrm{s}\Delta t_c}\ket{s_{k-1}, x_{k-1}, \sigma_{k-1}}\\
        && \times \prod_{k=\mathcal{N}+3}^{2\mathcal{N}+2} 
        \bra{s_k,x_k,\sigma_k}e^{iH_\mathrm{s}\Delta t_c^*}
        \ket{s_{k-1}, x_{k-1}, \sigma_{k-1}},
\end{eqnarray}
where
\begin{eqnarray}
	\bra{s_k,x_k,\sigma_k}e^{-iH_{\mathrm{s}}\Delta t_{\mathrm{c}}/\hbar}\ket{s_{k-1},x_{k-1},\sigma_{k-1}}=&&\\\nonumber
	\sum_{m=1}^{M_0} \phi_m(s_k,x_k,\sigma_k)\phi_m^*(s_{k-1},x_{k-1},\sigma_{k-1})&&
	e^{-iE_m\Delta t_{\mathrm{c}}/\hbar},
\label{eq:qm_prop}
\end{eqnarray}
$\phi_m$ and $E_m$ are the eigenstates 
and eigenenergies of $H_{\mathrm{s}}$, respectively, 
and $M_0$ is the number of eigenstates included in the expansion. 
The eigenstates and eigenvalues are obtained from a three-dimensional DVR grid calculation in terms of 
the solvent coordinate $\{-s^*, -s^*+\Delta s, \ldots, s^*-\Delta s, s^*\}$, 
the proton coordinate $\{-x^*, -x^*+~\Delta x, \ldots, x^*-\Delta x, x^*\}$, 
and the two electronic states. 

The discretized form of the non-local influence functional in 
Eq.~\ref{eq:cdef} is
\begin{equation}
	\mathcal{I}(\mathbf{s})=\mathcal{I}_0\;
	\text{exp}\left(-\sum_{k=1}^{2\mathcal{N}+2}\sum_{k'=1}^k 
	B_{kk'}s_k\;s_{k'}\right),
	\label{eq:bathinf}
\end{equation}
where  $\mathcal{I}_0$ is the partition function of the uncoupled bath oscillators \cite{rpf63, nm92, es01}.
For the case of linear system-bath coupling, the diagonal matrix elements are given by~\cite{nm92}
\begin{eqnarray}
	B_{kk}&&=\sum_{j=1}^f \frac{c_j^2}{M_j\omega_j^3
	\sinh(\beta\omega_j/2)} \nonumber\sin\left(\frac{\omega_j(t_{k+1}-t_{k})}{2}\right)\\
	&& \times\sin\left(\frac{\omega_j(t_{k+1}-t_{k}+i\beta)}{2}\right),
	\label{eq:bmat1}	
\end{eqnarray}	
and the off-diagonal matrix elements are given by
\begin{eqnarray}
	B_{kk'}&&=\sum_{j=1}^f \frac{c_j^2}{M_j\omega_j^3 
	\sinh(\beta\omega_j/2)} 	\sin\left(\frac{\omega_j(t_{k+1}-t_{k})}{2}\right) \nonumber \\
       &&\times\cos\left(\frac{\omega_j(t_{k+1}-t_{k'+1}+t_{k}-t_{k'}+i\beta)}{2}\right) \nonumber \\
	&&\times\sin\left(\frac{\omega_j(t_{k'+1}-t_{k'})}{2}\right).
	\label{eq:bmat2}
\end{eqnarray}
The complex times $t_k$ in Eqs. (\ref{eq:bmat1}) and (\ref{eq:bmat2}) are provided in Table~\ref{tab:quapi_times}.

\begin{table}[htb]
\caption{Complex times $t_k$  used to calculate the $\left\{B_{kk'}\right\}$.
\label{tab:quapi_times}}
	\begin{tabular}{cc}\toprule
\multicolumn{1}{c}{$k$}&\multicolumn{1}{c}{$t_k$}\\
\hline
$1$&$0$\\
$2, \ldots, \mathcal{N}+1$&$(k-1/2)\Delta t_\mathrm{c}$\\
$\mathcal{N}+2$&$t-i\beta\hbar/2$\\
$\mathcal{N}+3, \ldots, 2\mathcal{N}+2$&$(2\mathcal{N}+3/2-k)\Delta t_\mathrm{c}^*-i\beta\hbar$\\
$2\mathcal{N}+3$&$-i\beta\hbar$\\
\botrule
	\end{tabular}
\end{table}

The PCET flux operators, which correspond to the subset
projection operators defined in 
Eqs.\@ (\ref{eq:react_proj})-(\ref{eq:int_proj}), are
\begin{eqnarray}
\label{eq:fluxop1}
F_\text{p} &=& F_\mathrm{x}\ket{2}\bra{2} + h(x)F_\mathrm{e},\\
\label{eq:fluxop2}
F_\text{r} &=& -F_\mathrm{x}\ket{1}\bra{1} - h(-x)F_\mathrm{e},\\
\label{eq:fluxop3}
F_\text{i} &=& F_\mathrm{x}\left(\ket{1}\bra{1}-\ket{2}\bra{2}\right)
   + \left(h(-x)-h(x)\right) F_\mathrm{e},
\end{eqnarray}
where $F_\mathrm{x}~=~\frac{i}{\hbar}[H,h(x)]$ is the flux operator for the 
proton coordinate and 
$F_\mathrm{e}~=~\frac{i}{\hbar}[H,\ket{2}\bra{2}]$ is the flux operator
for the electronic diabatic states. 
The matrix elements appearing in Eq. (\ref{eq:fluxop1}) are evaluated using
\begin{eqnarray}
	\nonumber
&&\bra{s_k,x_k,\sigma_k}F_\mathrm{x}\ket{2}
	\braket{2}{s_{k'},x_{k'},\sigma_{k'}}=\\ 
&&\hspace{0.4in}\frac{i\hbar}{2m_x\; x_\text{FD}}\;\delta_{\sigma_k,2}\;
\delta_{\sigma_{k'},2}\;\delta(s_k-s_{k'})\\
	\nonumber
&&\hspace{0.4in}\times\left[ \delta(x_k)\delta(x_{k'}-x_\text{FD})-
\delta(x_k-x_\text{FD})\delta(x_{k'}) \right],
\label{eq:pflux_mat}
\end{eqnarray}
where $x_\text{FD}=\Delta x$, 
and 
\begin{eqnarray}
\nonumber
&&\bra{s_k,x_k,\sigma_k}h(x)F_\mathrm{e}\ket{s_{k'},x_{k'},\sigma_{k'}}=\\
\nonumber
&&\hspace{0.6in}\frac{i}{\hbar}h(x_k)\delta(x_k-x_{k'})V_{12}(s_k)\delta(s_k-s_{k'})\\
&&\hspace{0.6in}\times\left[\;\delta_{\sigma_k,1}\;\delta_{\sigma_{k'},2}
-\delta_{\sigma_k,2}\;\delta_{\sigma_{k'},1}\;\right],
\end{eqnarray}
respectively.
Similar expressions are used for the terms in 
Eqs. (\ref{eq:fluxop2}) and (\ref{eq:fluxop3}).

All convergence parameters for the QUAPI  calculations are provided in Table \ref{table:quapi_params}.
Convergence with respect to the number of path-integral beads is determined by comparing FAC functions with $\mathcal{N}=3-5$; the employed value of  $\mathcal{N}=4$ is consistent with the number for beads required in previous QUAPI simulations for proton and electron transfer reactions \cite{mt93a,mt96,es01}. 
Convergence with respect to the eigenfunction expansion is 
determined by comparing the trace over the complex-time propagator with $M_0=100-2000$; no significant numerical changes are found for cutoffs larger than those reported in Table \ref{table:quapi_params}. 
%

\begin{table}[htb]
	\caption{Convergence parameters for QUAPI.}
\label{table:quapi_params}
	\begin{tabular}{cccccccc}\toprule
\multicolumn{1}{c}{Model}&\multicolumn{1}{c}{$M_0$}&
\multicolumn{1}{c}{MC1$^{\rm a}$}&\multicolumn{1}{c}{MC2$^{\rm a}$}&
\multicolumn{1}{c}{$s^*$}&\multicolumn{1}{c}{$\Delta s$}&
\multicolumn{1}{c}{$x^*$}&\multicolumn{1}{c}{$\Delta x$}\\
\hline\\
I & 500 & $1\times10^7$ & $1\times10^8$ & $5$ & $0.1$ & $4$ & $0.2$ \\
II & 750 & $1\times10^8$ & $5\times10^8$ & $11$ & $0.09$ & $1$ & $0.14$\\
III & 500 & $1\times10^7$ & $1\times10^8$ & $5$ & $0.1$ & $4$ & $0.2$\\
\botrule
	\end{tabular}
\footnotetext{Indicates the number of Metropolis acceptance/rejection steps.}
\end{table}

For numerical comparison, the rate of the concerted PCET 
reaction in model I is calculated using both the QUAPI method
and a golden-rule rate expression. 
The QUAPI rate is obtained using
\begin{equation}
	k_\text{QUAPI}=\frac{k_\text{rp}}{Q_\text{r}},
	\label{eq:quapi_rate}
\end{equation}
where $k_\text{rp}$ is defined in 
Eq. (\ref{eq:def_notrate}) and $Q_\text{r}$ is the 
partition function for reactant subset $\Omega_r$; this 
 yields a rate $k_\text{Q}~=~1.7(1)~\times~10^{-6}$ a.u.
The golden-rule approximation for vibronically nonadiabatic PCET 
is given by \cite{ju79,db96,cc07,cv08,shs10}
\begin{eqnarray}
	\nonumber
	k_\text{GR}&=&\sum_\mu P_\mu \sum_\nu \frac{V_{\mu\nu}^2}{\hbar}
	\left(\frac{\beta\pi}{\lambda}\right)^{1/2}\\
	&&\hspace{0.2in}\times\text{exp}\left[\frac{-\beta(\Delta G^0+\lambda+\epsilon_\nu-\epsilon_\mu)^2}{4\lambda}\right],
	\label{eq:fgr_rate}
\end{eqnarray}
where $\epsilon_{\mu}$ and $\epsilon_\nu$ 
are the energies of the proton vibrational states $\mu$ and $\nu$, respectively, and  
$P_\mu$ is the thermal probability corresponding vibrational state $\mu$.
The PCET reaction in model I is ground-state dominated, electronically adiabatic and 
vibrationally nonadiabatic; construction of the diabats for 
model I thus yields a vibronic coupling of $V_{00}=3.75\times10^{-4}$,
solvent reorganization energy 
$\lambda=1.34\times10^{-2}$ ~a.u., and driving force $\Delta G^0=0$.  
The golden-rule rate obtained for the concerted reaction in model I 
is $k_\text{GR}~=~2.06~\times~10^{-6}$ a.u.,
which is in good agreement with the QUAPI rate stated above.


\newpage

\end{document}